\documentclass[12pt]{article}
\usepackage{latexsym}
\usepackage{amssymb}
\usepackage{epsf}
\usepackage{amsmath}
\usepackage[hypertex]{hyperref}
\usepackage{graphicx}% Include figure files

\newcommand{\gsim}{\gtrsim}

\begin{document}
\pagestyle{empty}

\begin{flushright}
%\today 
\end{flushright}

\hfill TU-862\\
\vspace{1cm}

\begin{center}

\vspace{1.5cm}
{\Large\bf 
%Is the Dark Matter in the Minimal Model or not?
Constraints on Scalar Phantoms
}

\vspace{1cm}
{\large Masaki Asano and Ryuichiro Kitano}

\vspace{1cm}

{\it {Department of Physics, Tohoku University,
    Sendai 980-8578, Japan}}

\vspace{1cm}
\abstract{ We update the constraints on the minimal model of dark
matter, where a stable real scalar field is added to the standard model
Lagrangian with a renormalizable coupling to the Higgs field.
Once we fix the dark matter abundance, there are only two relevant model
 parameters, the mass of the scalar field and that of the Higgs boson.
The recent data from the CDMS~II experiment have excluded a parameter
 region where the scalar field is light such as less than about 50~GeV.
In a large parameter region, the consistency of the model can be tested
 by the combination of future direct detection experiments and the LHC
 experiments.

 }

\end{center}

%\renewcommand{\theequation}{\thesection.\arabic{equation}}
%\renewcommand{\thepage}{\arabic{page}}
%\setcounter{page}{1}
%\renewcommand{\thefootnote}{\#\arabic{footnote}}
%\setcounter{footnote}{0}

%%%%%%%%%%%%%%%%%%%%%%%%%%%%%%%%%%%%%%%%%%%%%%%%%%%%%%%%%%%%%%%%%%%%%%%%%%%%
\newpage
\baselineskip=18pt
\setcounter{page}{2}
\pagestyle{plain}

\setcounter{footnote}{0}

Weakly interacting massive particle (WIMP) is one of the most attractive
scenarios to explain the dark matter component of the universe. The
simplest and the most economical model to realize this scenario is the
model of the scalar dark matter, where a new gauge singlet real scalar
field $S$ is introduced to the standard model and it has an interaction
with the Higgs field, $H$. The following Lagrangian density is added to
the standard model:
\begin{eqnarray}
 {\cal L}_{S} = {1 \over 2} \partial_\mu S \partial^\mu S
-{m^2 \over 2} S^2 - {k \over 2} S^2 |H|^2 - {h \over
  4!} S^4.
\end{eqnarray}
This model has been first proposed in Ref.~\cite{Silveira:1985rk} 
(it is called scalar phantoms)
and further studied in
Refs.~\cite{McDonald:1993ex,Burgess:2000yq,McDonald:2001vt,Davoudiasl:2004be,
Cynolter:2004cq,Barger:2007im,He:2008qm,Gonderinger:2009jp}.
One can obtain the correct size of the dark matter density via the
standard thermal decoupling process in the early universe.

The model can be either thought of as a device to accommodate dark matter
in new physics models or as a fundamental theory to describe the physics
up to the Planck scale.
If we take the latter attitude, the parameter region for $k$ and $h$ is
subject to the constraint from perturbativity and stability of the
potential~\cite{McDonald:2001vt,Davoudiasl:2004be,Cynolter:2004cq,Gonderinger:2009jp}.

Because the model is compact, there are only four unknown parameters:
the Higgs boson mass $m_h^2$, the mass of dark matter $m_S^2 (\equiv m^2
+ k \langle H \rangle^2 = m^2 + k v ^2)$, and two coupling constants $k$
and $h$. Here $v \simeq 174 {\rm GeV}$. Since the coupling constant of 
the dark matter self-interaction $h$ is not important for most of 
interesting physical observables, there are only three relevant parameters, 
one of which can be fixed by requiring the correct dark matter abundance.
Therefore, the model is highly predictive. The parameter space is a
two-dimensional $m_h - m_S$ plane.

In this article, we update the constraints on the model in light of new
data from the CDMS II experiment~\cite{Ahmed:2009zw} as well as the
Tevatron exclusion~\cite{Collaboration:2009je} of a Higgs boson mass region.
We discuss the interplay between the dark matter detection experiments and
the Higgs boson searches at the LHC experiments.\footnote{Similar studies have been done in Refs.~\cite{He:2009yd,Farina:2009ez} very recently.}

The scalar field $S$ is stable due to a discrete symmetry $S
\leftrightarrow -S$, and therefore is a candidate for a non-baryonic cold
dark matter.
In the WIMP scenario, the number density of $S$ is determined by the
annihilation cross section, 
$\langle\sigma_{\rm ann.}v\rangle$, which is proportional to $k^2$.
One can fix the coupling constant $k$ by requiring the abundance to
explain the dark matter component of the energy density of the
universe~\cite{Hinshaw:2008kr}.
The energy density of dark matter is given by
\begin{eqnarray}
 \Omega_S \simeq {1.8 \times 10^{-10}~{\rm GeV}^{-2} \over \langle\sigma_{\rm ann.}v\rangle},
\end{eqnarray}
The annihilation cross section is given
by
\begin{eqnarray}
 \langle\sigma_{\rm ann.}v\rangle = {4 k^2 v^2 \over 
(4 m_S^2 - m_h^2)^2 + m_h^2 \Gamma_H^2
}
\cdot
{\Gamma_H|_{m_h = 2 m_S} \over 2 m_S}
\end{eqnarray}
for $m_S < m_h$, and 
\begin{eqnarray}
  \langle\sigma_{\rm ann.}v\rangle = {4 k^2 v^2 \over 
(4 m_S^2 - m_h^2)^2 + m_h^2 \Gamma_H^2
}
\cdot
{\Gamma_H|_{m_h = 2 m_S} \over 2 m_S}
+
{k^2 \sqrt{1 - {m_h^2 / m_S^2}} \over 64 \pi m_S^2} 
\left(
1 + {3 m_h^2 \over 4 m_S^2 - m_h^2}
\right),
\end{eqnarray}
for $m_S \geq m_h$.
Here $\Gamma_H$ and $\Gamma_H|_{m_h = 2m_S}$ 
are the total decay width
of the Higgs boson and that with a mass $2 m_S$, respectively.
(One should take out $h \rightarrow SS$ mode in $\Gamma_H|_{m_h = 2m_S}$.)
By using the WMAP data~\cite{Hinshaw:2008kr}, $\Omega_S \simeq 0.11$, 
one can fix $k$.

Dark matter $S$ interacts with nuclei through an exchange of
the Higgs boson.
Direct detection experiments such as CDMS II can therefore put 
constrains on the model.
The spin-independent cross section of the $S$-nucleus elastic scattering
normalized to a nucleon is given by
\begin{eqnarray}
 \sigma_{SI} = {1 \over 4 \pi} {(1~{\rm GeV})^2 \over A^2 m_S^2} (Z f_p + (A-Z)
  f_n)^2,
\label{eq:W-N_CS}
\end{eqnarray}
where $A$ and $Z$ are mass and atomic numbers of the target nucleus,
respectively. 
The $f_p$ and $f_n$ factors are given by
\begin{eqnarray}
f_{N} = \frac{k m_{N}}{m_h^2} 
\sum_{q} f_q^N, \ \ \ \mbox{and}\ \ \ 
f_q^N =  {m_q \langle N | \bar q q | N \rangle \over m_N},
\label{eq:fp}
\end{eqnarray}
where $q=u,d,s,c,b,t$ and $N=p,n$. 
In this study, we use 
\begin{eqnarray}
 f_u^p \!\!\!&=&\!\!\! 0.021, \quad  f_d^p = 0.029, \quad f_s^p = 0.0, \\
 f_u^n \!\!\!&=&\!\!\! 0.016, \quad  f_d^n = 0.037, \quad f_s^n = 0.0,\\
 f_c^p \!\!\!&=&\!\!\! f_b^p = f_t^p = f_c^n = f_b^n = f_t^n = 0.070,
\end{eqnarray}
which give a conservative estimate of the scattering cross section~\cite{Corsetti:2000yq,Ohki:2008ff,Cheng:1988im,Ohki:2009mt}.

The Higgs boson can decay into a pair of $S$ if $ m_S < m_h / 2 $. This
invisible decay width is given by:
\begin{eqnarray}
 \Gamma_{h \rightarrow SS} = \frac{k^2 v^2}{16 \pi m_h}
                                 \sqrt{ 1 - \frac{4 m_S^2}{m_h^2}}.
\label{eq:Invisible decay width}
\end{eqnarray}
The strategy to search for the Higgs boson in collider experiments will
be tightly connected to dark matter physics in this model.

There have been studies of the invisible Higgs decay at the LHC.
The discovery potential is a function of 
\begin{eqnarray}
 \xi^2 = 
{\sigma_{\rm BSM} \over \sigma_{\rm SM}}
\cdot B_{\rm inv}, 
\end{eqnarray}
where $B_{\rm inv}$ is the branching fraction of the invisible decay
mode, and $\sigma_{\rm BSM}$ and $\sigma_{\rm SM}$ are the production
cross sections of the Higgs boson in new physics models and in the
standard model, respectively.
At the ATLAS detector with $30~{\rm fb^{-1}}$, it is possible to
discover the invisible decay of the Higgs boson in the $100$ GeV $< m_h <
200$ GeV region if $\xi^2 \gsim 60~\% $~\cite{Aad:2009wy}. 
In the present model, $\xi^2 = B_{\rm inv}$ because the production cross
section is same as the one in the standard model.

\begin{figure}[t]
\begin{center}
\begin{tabular}{c}
\includegraphics[scale=0.5, angle=0]{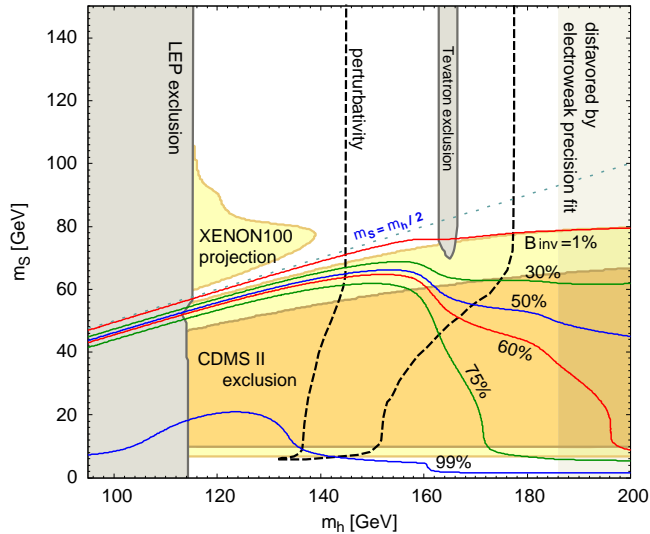}
\end{tabular}\\
\caption{\label{fig:Direct} Experimental constraints and contours of the
branching ratio of the invisible Higgs boson decay. The region inside
the dashed lines satisfies the stability and triviality bounds for
$h(m_Z) = 0$.}
\end{center}
\end{figure}

\label{sec:Exbounds}
\setcounter{equation}{0}

Let us discuss the current experimental constraints on this model and
the size of $B_{\rm inv}$ which will be important for the Higgs boson
search.
As explained before, one can discuss the constraints/prediction of the
model on the $m_S - m_h$ plane by requiring the correct dark matter
abundance (Fig.~\ref{fig:Direct})\footnote{In calculating the dark
matter abundance and the decay width of the Higgs boson, we have used
the program HDECAY~\cite{Djouadi:1997yw}.}.
The recent data from the CDMS~II experiment excludes the region with
$10~{\rm GeV} \lesssim m_S \lesssim 50$~GeV although there are
uncertainties from the local density of dark matter. We also overlaid
the projected sensitivity of the XENON100 experiment~\cite{Aprile:2009yh}.
The bound from the Higgs boson searches are also shown.
The shaded region in the left side is excluded by the LEP-II direct
Higgs boson searches~\cite{Barate:2003sz,:2001xz}. 
The excluded region from the Tevatron experiments, $163$ GeV $< m_h <
166$ GeV~\cite{Collaboration:2009je}, is also shown where the bound
disappears for $m_S \lesssim 70$~GeV due to large $B_{\rm inv}$.
The region $m_h > 186$~GeV is disfavored by the electroweak precision
measurements (95\% CL)~\cite{Collaboration:2009jr}.
A very light dark matter region, $m_S \lesssim 1 {\rm GeV}$, is ruled
out by a large branching ratio of the $B \rightarrow K S S$
decay~\cite{Bento:2001yk,Bird:2004ts,Bird:2006jd,Kim:2009qc}.
It has been studied that a heavier region can be explored by 
the invisible decay of $\Upsilon$ (or other heavy meson)~\cite{Yeghiyan:2009xc}.
Inside the dashed line, the model is perturbative and the Higgs
potential is stable up to the Planck scale.
It is interesting to notice that the recent experimental results started
to explore a parameter region of the model.

We also show contours of $B_{\rm inv}$ (solid lines) on the same plane. 
In the dark matter detection and the Higgs search experiments, there are
in principle four physical observables, $m_h$, $B_{\rm inv}$, $m_S$, and
$\sigma_{SI}$, whereas in this model $B_{\rm inv}$ and $\sigma_{SI}$ can
be calculated in terms of other two, $m_h$ and $m_S$.
Therefore, one can check the consistency by using data from both
experiments.
Almost entire region in Fig.\ref{fig:Direct} will be covered by next
generation experiments such as SuperCDMS, XENON, and
XMASS~\cite{Aprile:2009yh,Abe:2008zzc,Brink:2005ej} except for a region very close to the $m_S = m_h
/2 $ line. Also, the Higgs boson search at the LHC will cover most of
the region either by the invisible mode or by the ordinary standard
model Higgs searches.

Finally, we comment on the case where two events reported by CDMS~II is
a signal of dark matter. 
In this case, the parameter region is restricted near the boundary of the
CDMS~II exclusion region.
Interestingly, the mass region preferred by the data, 
$m_S \lesssim 100 {\rm GeV}$ \cite{Kopp:2009qt,Farina:2009ez}
, is consistent with the prediction in Fig.~\ref{fig:Direct}.
Since $B_{\rm inv}$ is rather large in the region, the LHC experiments
should confirm the scenario by observing the invisible decay of the
Higgs boson when $m_h \lesssim 160$~GeV\footnote{It may be possible to
discover the invisible mode even for $m_h \gtrsim 160$~GeV, i.e., $\xi^2
\sim 30~\%$, once systematic uncertainties are better
understood~\cite{Aad:2009wy}.}.

The positron excess reported by the PAMELA experiment \cite{Adriani:2008zr} cannot be
simultaneously explained. One can assume that decays of dark matter
provide a source of high-energy positrons through a small breaking of
the $Z_2$ symmetry. In order to explain the spectrum, $m_S$ is required
to be larger than 200~GeV, that is not compatible with the explanation
of the CDMS~II data in this model.

\section*{Acknowledgments}
This work was supported in part by the Grant-in-Aid for the Global COE Program 
“Weaving Science Web beyond Particle-matter Hierarchy” from the Ministry of Education, 
Culture, Sports, Science and Technology of Japan (M. A.) and 
the Grant-in-Aid for Scientific Research from the Ministry of Education, Science, Sports, 
and Culture of Japan, no.\
21840006 (R.K.).

\end{document}